\documentclass[twocolumn]{aastex6}
\usepackage{graphicx}
\usepackage{amsmath,amssymb}
\usepackage{units}

\newcommand{\BF}{\begin{figure}\begin{center}}
\newcommand{\EF}{\end{center}\end{figure}}
\newcommand{\BE}{\begin{equation}}
\newcommand{\EE}{\end{equation}}
\newcommand{\BEA}{\begin{eqnarray}}
\newcommand{\EEA}{\end{eqnarray}}
\newcommand{\ti}{\textit}
\newcommand{\tr}{\textrm}

\newcommand{\ms}{M_{\odot}}

\shorttitle{Evidence for a Dusty Dark Dwarf Galaxy in MG\,0414+0534}
\shortauthors{Inoue, Matsushita, Minezaki and Chiba}

\begin{document}


\title{Evidence for a Dusty Dark Dwarf Galaxy 
in the Quadruple Lens MG\,0414+0534}

\author{Kaiki Taro Inoue}
\affiliation{Faculty of Science and Engineering, 
Kindai University, Higashi-Osaka, 577-8502, Japan}

\author{Satoki Matsushita}
\affiliation{Academia Sinica Institute of Astronomy and Astrophysics, 
P.O. Box 23-141, Taipei 10617, Taiwan, Republic of China}

\author{Takeo Minezaki}
\affiliation{Institute of Astronomy, School of Science, University of
Tokyo, Mitaka, Tokyo 181-0015, Japan}

\author{Masashi Chiba}
\affiliation{Astronomical Institute, Tohoku University,
Aoba-ku, Sendai 980-8578, Japan}

\begin{abstract}
We report the $4 \, \sigma$ detection of a faint object with a flux of
 $\sim 0.3\, \tr{mJy}$, in the vicinity of the 
quadruply lensed QSO MG\,0414+0534 
using the Atacama Large Millimeter/submillimeter array (ALMA) Band 7. 
The object is most probably a dusty dark dwarf galaxy, 
which has not been detected in either the optical,
 near-infrared (NIR) or radio (cm) bands.
An anomaly in the flux ratio of the lensed images observed in 
Band 7 and the mid-infrared (MIR) band and the reddening of the QSO light color
can be simultaneously explained if we consider the object as a lensing 
substructure with an ellipticity $\sim 0.7$ at a redshift of
$0.5 \lesssim z \lesssim 1$.
Using the best-fit lens models with three lenses, 
we find that the dark matter plus baryon mass associated with the object 
is $\sim 10^9\, \ms$, the dust mass is $\sim 10^7\,\ms$ and the
 linear size is $\gtrsim 5\,$kpc. Thus our findings
suggest that the object is a dusty dark dwarf galaxy. A 
substantial portion of faint submillimeter galaxies (SMGs) in the 
universe may be attributed to such dark objects.  
\end{abstract}
\section{Introduction} \label{sec:intro}
The flux ratios of lensed images in some quadruply lensed
QSOs disagree with the prediction of best-fit lens models with a smooth 
potential whose fluctuation scale is larger than the separation between the
lensed images. Such a discrepancy is called 
the ``anomalous flux ratio'' and 
has been considered as an imprint of cold dark matter subhalos
with a mass of $\sim 10^{8-9} \ms$ in the lensing
galaxy \citep{mao1998, chiba2002, dalal-kochanek2002, metcalf2002,
kochanek2004, metcalf2004, chiba2005, sugai2007, more2009, 
minezaki2009, xu2009, xu2010}. Subhalos can be also detected through 
their effects on the image positions 
\citep{ros2000, inoue2005a, inoue2005b, koopmans2005, vegetti2012, vegetti2014, hezaveh2016}.

However, the flux-ratio anomalies can be also explained by the
weak lensing effects due to intervening halos and voids with a mass
scale of $\lesssim 10^{9} \ms$ in the line-of-sight \citep{metcalf2005a, xu2012, inoue-takahashi2012, takahashi-inoue2014,inoue2015, inoue-minezaki2016}. 
For lens systems with a source at a 
high redshift $z\gtrsim 2$, \citet{inoue2016} has recently argued that 
the main cause of the anomaly is structures in the intervening 
line-of-sights rather than subhalos associated with 
the lensing galaxy. In order to measure the 
redshift of possible perturbers, we need to observe
the flux ratios and positions of lensed images as precisely as possible.

In this paper, we present our result on Atacama 
Large Millimeter/submillimeter Array (ALMA) 
continuum observations of the quadruply lensed,
radio-loud QSO MG\,0414+0534.   
It has been known that MG\,0414+0534 shows a strong sign
of anomaly in the flux ratio and reddening
in the optical and near-infrared (NIR) band (see 
\citealt{minezaki2009} and references therein). The origin of these
features has not been fully understood yet. 
Throughout this paper, we use the Planck 2016 cosmological parameters:
$h=0.678,\Omega_m=0.308$ and $\Omega_\Lambda=0.692$ \citep{planck2016}.  

\section{Observations and data reduction} 
The observations of MG\,0414+0534 were 
carried out with ALMA at Band 7 on 2015
June 13 and August 14 as part of Cycle 2 
(Project ID: 2013.1.01110.S, PI: K.T. Inoue).
The number of antennas used in the observation in June was 35
and that in August was 42. The phase center was 
$\alpha=$04$^\tr{h}$14$^\tr{m}$37$^\tr{s}$.7686, $\delta=$+05$^\circ$34$'$42\farcs 352 (J2000).
The total on-source integrating times were
104.63 minutes and 52.28 minutes and the recoverable 
largest angular size was $\sim 3.3$\,arcsec.
The ALMA correlator was configured 
to have four spectral windows centered at
335.0\,GHz, 337.0\,GHz, 347.0\,GHz and 349.0\,GHz with a
bandwidth 2\,GHz and a frequency width of 15.625\,MHz for each window.

The calibration and imaging of the data were carried out using the
Common Astronomy Software Application package (CASA;
\citealt{mcmullin2007}). 
The continuum images produced with a natural or a robust ($\ti{robust}=-1$) 
weighting of the visibilities were used for
imaging and modeling, respectively. We used 
$\ti{multi-scale}$ imaging \citep{cornwell2008} 
with scales of 0, 6, 12, 24 and 48 for the natural weighting 
(pixel size is 0.02\,arcsec) and
0, 6, 12 and 24 for the robust ($\ti{robust}=-1$) weighting 
(pixel size is 0.01\,arcsec). The images have not 
been corrected for the primary beam attenuation because 
the source is small and located at the center of the primary beam.
The 1$\,\sigma$ noises in the flux density per pixel 
are 49\,$\mu$Jy per beam and 140\,$\mu$Jy per beam, 
the full widths at half maximum of the elliptical gaussian 
fitted to the dirty beam are ($0\farcs278\times 0\farcs243$)
and ($0\farcs135\times 0\farcs111$) and the PA's measured in 
degrees East of North are $46\fdg5$ and $75\fdg4$, for the natural
and robust ($\ti{robust}=-1$) weightings, respectively.
\section{Results}
\subsection{Dust continuum emission}
\begin{figure*}
\epsscale{1.1}
\hspace{-1cm}
\plotone{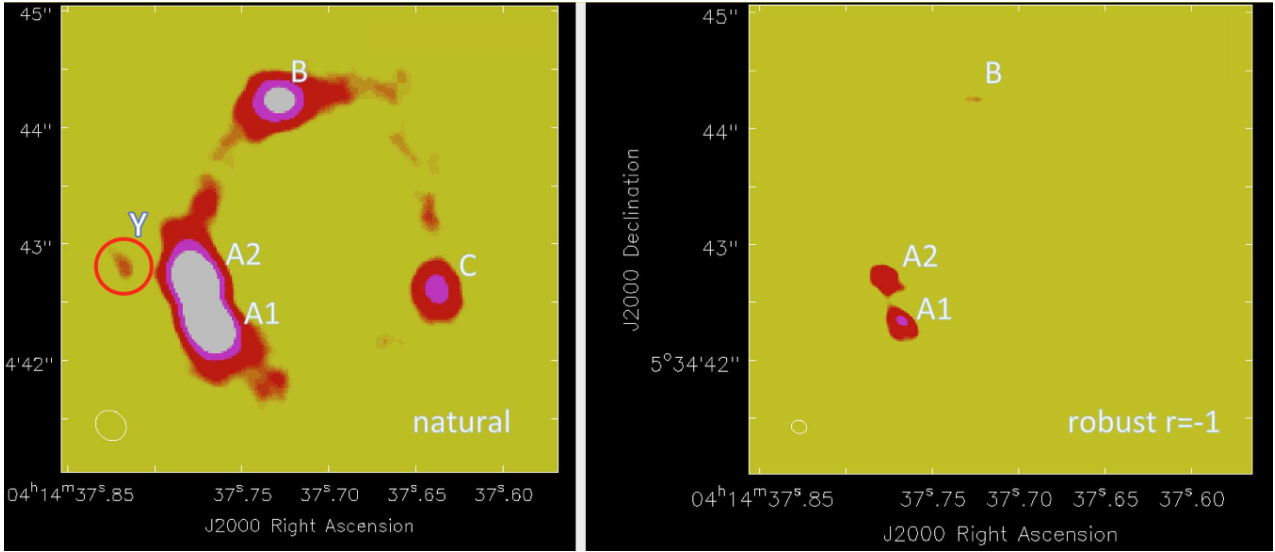}
\label{f1}
\caption{ALMA $0.88\,\tr{mm}$ (Band 7) dust continuum images of MG\,0414+0534. The
 surface brightness ($\gtrsim 4\, \sigma$) is plotted in color
for a natural weighting (left) and a 
robust ($\ti{robust}=-1$) weighting (right). The 
sizes of a Gaussian fitted beam, which are shown in the 
bottom-left corner of each image are $278\times 243\,\tr{mas}$ and 
$135\times 111\,\tr{mas}$, respectively. A faint spot inside a red circle
 (left) is an object Y. }
\end{figure*}
The dust continuum \citep{barvainis2002} image
obtained with a natural weighting 
(Figure 1, left) shows bright quadruply lensed spots A1, A2, B and C 
and a faint Einstein ring. As shown, lensed spots A1, A2 and B are resolved as separate regions if a robust $r=-1$ weighting is used. The aperture flux
ratios\footnote{Within an aperture radius of $0.21$\,arcsec, the 
neighboring pixels with flux density $>3.5\,
\sigma$ are all included.} for a robust $r=-1$ weighting 
were $(\tr{A2}/\tr{A1}, \tr{B}/\tr{A1})
=(0.87\pm 0.03,0.38\pm 0.06)$. Note that the errors  
were estimated using the flux density in apertures of the same size
placed at source-free locations. We assumed that correlations 
between errors are negligible. 
The obtained result is consistent with the flux ratios
in the MIR band $(\tr{A2}/\tr{A1}, \tr{B}/\tr{A1}, \tr{C}/\tr{A1})
=(0.919\pm 0.021,0.347\pm 0.013, 0.139\pm 0.014)$ \citep{minezaki2009,
macleod2013}.

Interestingly, the image obtained with a natural weighting 
showed a faint object in the 
vicinity of A2, which we call 'object Y' (shown in a red circle in 
Figure 1, left). The apparent angular size is equal to or smaller than 
the beam size. The flux density of Y inside an aperture radius of $0.26\,\tr{arcsec}$ centered at the peak is $0.26\,\tr{mJy}$ and the locally maximum value for
the aperture flux is $0.27\,\tr{mJy}$. The corresponding statistical significances, which were estimated using the flux density in apertures with the same size
placed at source-free locations, were $4.0\,\sigma$ and $4.1\,\sigma$,
respectively. A gaussian fit to the intensity of the faint emission
yielded a peak flux density of $0.2\, \tr{mJy}/\tr{beam}$ corresponding
to $4.0\, \sigma$.

Assuming a dust mass opacity coefficient, $\kappa_{850\mu \tr{m}/(1+z_{\tr{Y}})}
=0.077\,(1+z_{\tr{Y}})^\beta\tr{m}^2\tr{kg}^{-1}$ \citep{dunne2000}, for 
an emissivity index $\beta=1-2$ with 
a dust temperature $T_\tr{d}=20-50\,\tr{K}$ 
and a dust redshift $z_{\tr{Y}}=0.5-1$ (see subsection 3.2), 
the dust mass is $M_\tr{d}=10^{6-7}\,\ms$. In the following, we
interpret Y as the central core of a dusty dwarf galaxy.
\begin{figure*}
\epsscale{1.2}
\hspace{-0.2cm}
\plotone{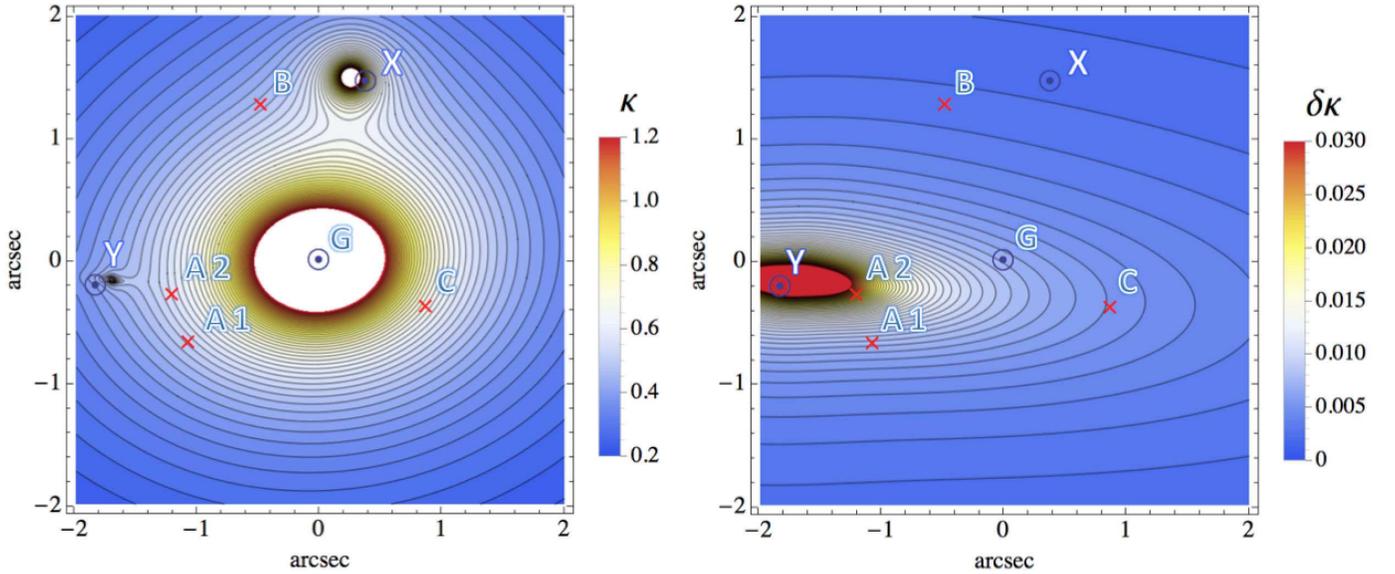}
\label{f2}
\caption{The total convergence $\kappa$ (left) and the 
convergence contribution $\delta \kappa $ from object Y (right) 
for a best-fit lens model with $e_\tr{Y}=0.72$ and $z_{\tr{Y}}=0.548$. 
Circled dots represent the positions of the observed centroids of the primary 
lensing galaxy G, object X and object Y. Red crosses represent the
 position of the lensed HST images of MG\,0414+0534
 \citep{falco1997}. The surface mass density at A2 is approximately three times larger
 than that at A1. }
\end{figure*}
\begin{deluxetable*}{lccc}
\tablecaption{Modeling Results for MG\,0414+0534}
\tablewidth{0pt}
\tablehead{Model & \colhead{\!\!G+X} &                    
\colhead{\!\!\!\!\!G+X+Y(subhalo)}&                    
\colhead{\!\!\!\!\!G+X+Y(line-of-sight)}
}                                                                               
\startdata 
$b_{\textrm{G}}('')$ & 1.104 & 1.113 &   1.114   \\
$(x_s,y_s)('')$ & (-0.0663, 0.2603)& (-0.0919, 0.2130) & (-0.0950, 0.2157)  \\
$e_{\tr{G}}$ & 0.303&0.223 & 0.225  \\
$\phi_{\tr{G}}$(deg)&  -87.9 &-85.4 &  -85.4  \\
$\gamma$ & 0.0878 &0.0848 & 0.0862 \\
$\phi_\gamma$(deg) & 47.5 &54.7 &  54.7 \\
$(x_{\tr{G}}, y_{\tr{G}})('')$ & (0.0020, -0.0011) &(0.0006, -0.0001)&(-0.0003, 0.0006)    \\
$b_\textrm{X}('')$&0.199  &0.154 &0.154   \\
$(x_{\tr{X}}, y_{\tr{X}})('')$ &(0.390, 1.513)&(0.274, 1.457)& (0.256, 1.495)   \\
$r_{\tr{X}}('')$ & 0.007 & 0.032 & 0.033 \\
$b_{\tr{Y}}('')$ & &0.008 &0.015 \\
$(x_{\tr{Y}}, y_{\tr{Y}})('')$&  &(-1.751, -0.1151)&(-1.692, -0.152)\\
$e_{\tr{Y}}$&  &0.72  & 0.72\\
$\phi_{\tr{Y}}(\tr{deg})$ & &85.74 & 85.73  \\
$z_{\tr{Y}}$& &0.958 & 0.548 \\
$\chi^2_{\tr{pos}}$ &6.35  & 1.42 &2.50 \\
$\chi^2_{\tr{flux}}$& 22.09 & 3.41 &  2.93\\
$\chi^2_{\tr{weak}}$& 0.57  &0.57 &0.57  \\
$\chi^2_{\tr{tot}}/$dof &29.0/4 &5.4/2 & 6.0/2 \\
A2/A1 &1.005 & 0.925  & 0.932 \\
B/A1 &0.342  & 0.350  & 0.351 \\
C/A1 &0.171  & 0.164  & 0.161 \\
\enddata
\tablecomments{$\chi^2_{\tr{tot}}$ is the sum of contributions from the
image and lens positions $\chi^2_{\tr{pos}}$, the flux ratios
$\chi^2_{\tr{flux}}$, and the weak priors $\chi^2_{\tr{weak}}$ 
on $e_{\tr{G}}$, $\gamma$ and $e_{\tr{Y}}$ if applicable. The coordinates are
centered at the centroid of the primary lensing galaxy G (CASTLES
 database) \citep{falco1997}.}
\end{deluxetable*}
\begin{figure*}[t]
\epsscale{1.2}
\hspace{-0.5cm}
\plotone{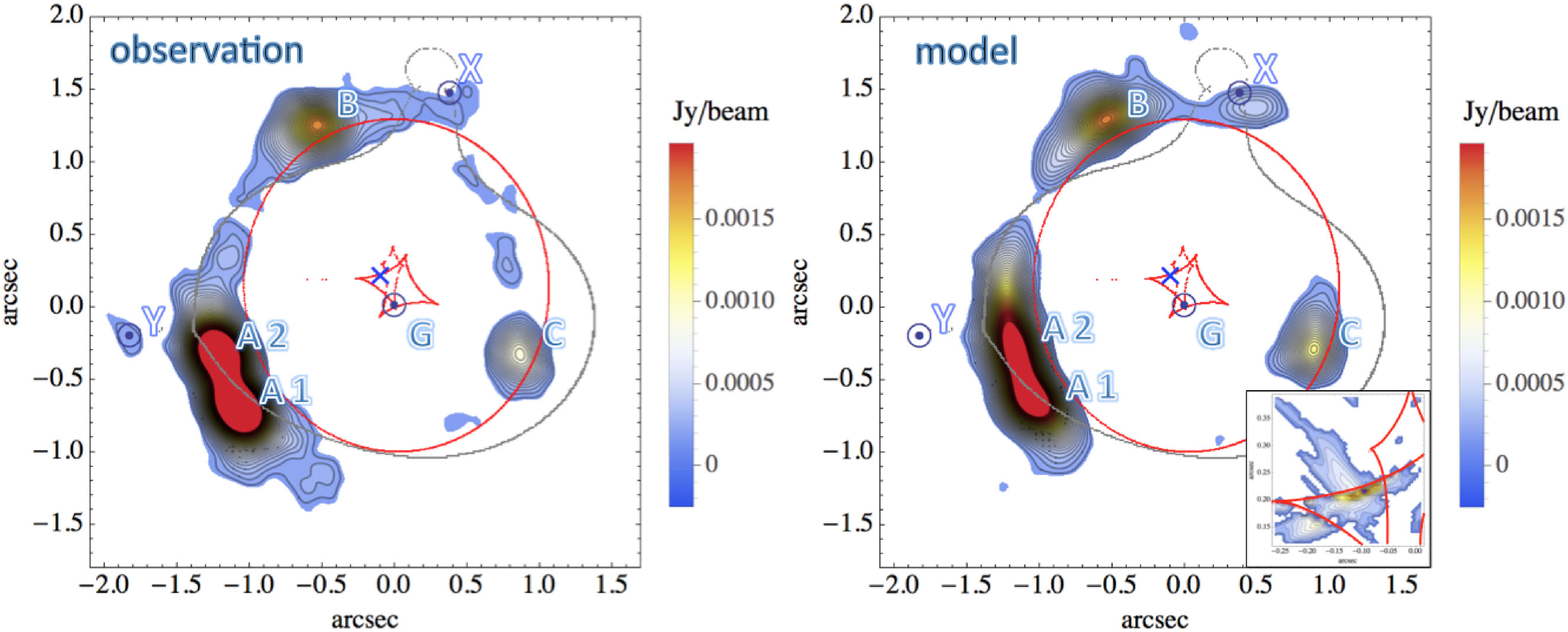}
\label{f3}
\caption{Observed and modeled dust continuum images for a three-galaxy model (G+X+Y) 
with $e_\tr{Y}=0.72$ and $z_{\tr{Y}}=0.548$ (see Table 1). The both
 images were produced with a natural weighting. The color shows
the surface brightness ($> 3.5\, \sigma$) and the contour spacing is
$1\,\sigma$. The inset in the right panel shows the source model.  
The red curves are the caustics and cuts and the gray curves are the critical curves. The circled dots are the
 positions of the observed centroids of G, X and Y and the 
blue crosses represent the position of the QSO core in the source plane.}
\end{figure*}
\subsection{Flux-Ratio Anomaly}
The source of MG\,0414+0534, at a redshift of $z_{\tr{S}}=2.639$, is lensed by a
foreground elliptical galaxy G (the primary lens) at $z_{\tr{L}}=0.9584$ 
\citep{hewitt1992, lawrence1995, tonry1999}. It has been known 
that a canonical model, a singular isothermal ellipsoid plus an external
shear (SIE-ES) does not give a good fit. 
In order to improve the fit, we need to  
take into account a possible satellite galaxy, object X, which 
was detected in the I band HST/NICMOS image \citep{schechter1993,
ros2000, trotter2000}. However, the observed mid-infrared (MIR) flux ratio 
of A1 to A2 is still not consistent with the SIE-ES-X model prediction even if 
low-multipole modes $m=3,4$ are added to the potential. If a subhalo
resides in the vicinity of A2, the anomalous flux ratio A1/A2
can be explained \citep{minezaki2009, macleod2013}. 

First, we used a canonical model, an SIE-ES for G 
plus a cored singular isothermal sphere (cored-SIS) for X. 
The model has 13 parameters: the position $(x_{\tr{G}}, y_{\tr{G}})$, 
effective Einstein angular radius $b_\tr{G}$\footnote{We took a
definition of $b_G$ for which the mass inside $b_G$ coincides
with that inside the critical curve \citep{kormann1994}. }, 
ellipticity $e_{\tr{G}}$ and position angle $\phi_{\tr{G}}$ for G, 
the position $(x_{\tr{X}}, y_{\tr{X}})$, 
effective Einstein angular radius $b_\tr{X}$, core radius
$r_{\tr{X}}$ for X, the amplitude $\gamma$ 
and position angle $\phi_{\gamma}$ (direction of the associated mass
clump) of the external shear and the source
position $(x_s,y_s)('')$. All the position
angles are measured in degrees East of North. 
The redshift $z_\tr{X}$ of X was assumed to be $z_\tr{X}=z_{\tr{L}}$ 
\citep{macleod2013}\footnote{We found that a deviation of $z_\tr{X}$ from
$z_{\tr{L}}$ yields a worse fit.}. We used the measured HST WFPC/WFPC2 
positions of lensed images and centroids of 
lensing galaxies, G and X in the CASTLES data
archive\footnote{http://www.cfa.harvard.edu/castles/} \citep{falco1997},
and the MIR flux ratios, A2/A1 and B/A1 \citep{minezaki2009,
macleod2013}. Note that MIR and submillimeter fluxes are free 
from microlensing by stars and dust extinction whereas optical and NIR fluxes are
potentially influenced by these systematics.

We assumed a weak prior for the amplitude of the external shear
$\gamma(=0.05\pm0.05)$ and the ellipticity $e_{\tr{G}}(=0\pm 1)$. We also
assumed an error of 0.2\,arcsec for the position of the centroid of 
X, which is roughly the angular size, and 0.003\,arcsec for the position
of the centroid of G. A large error for X was adopted because of possible
contamination from the lensed fifth image. We had 
$\chi_{\tr{tot}}^2/\tr{dof}=29.0/4$ 
with a best-fit flux ratio $\tr{A}2/\tr{A}1=1.005$, 
which implies that the fit is
not satisfactory (see Table 1 for the definition of $\chi_{\tr{tot}}^2$). 

Next, we added an SIE for modeling Y. The
parameters are the position $(x_{\tr{Y}}, y_{\tr{Y}})$, 
effective Einstein angular radius $b_\tr{Y}$, ellipticity $e_{\tr{Y}}$,  
and position angle $\phi_{\tr{Y}}$. As a first guess,
we assumed $z_{\tr{Y}}=z_{\tr{L}}$. The astrometric error of Y is assumed to be
0.13\,arcsec, which is roughly the angular size. 
We obtained the coordinates of the centroid of Y ($-1\farcs824,-0\farcs212$)
by matching the brightest spots in the ALMA and CASTLES data. The residual errors were $\sim 0.01\,$arcsec.

Then, it turned out that the fit was greatly improved provided that the
ellipticity $e_{\tr {Y}}$ of Y satisfies $e_{\tr {Y}}\gtrsim 0.7$. We
obtained a better fit for a larger ellipticity $e_{\tr{Y}}$.
For explaining the reddening of A1 and A2 by dust extinction, however, very 
large ellipticity is not likely. Therefore, 
we assumed another weak prior on the ellipticity
of Y as $e_{\tr{Y}}=0\pm 1$, which results in 20 total constraints.  
For $e_{\tr{Y}}=0.72$, we had $\chi_{\tr{tot}}^2/\tr{dof}=5.4/2$ 
with a best-fit flux ratio A2/A1$=0.925$ (Table 1). For models with a larger
ellipticity, the ratio of convergence of Y at A1 and A2 becomes smaller
than $\sim 1/3$, and hence the convergence 
at A1 becomes too small to account for
the reddening in optical/IR bands (see subsection 3.3).

We also examined the dependence of the fit on the
redshift $z_\tr{Y}$ of Y. We found that $\chi_{\tr{tot}}^2$ is nearly 
constant if $0.5 \lesssim z_{\tr{Y}} \lesssim z_{\tr{L}}$. 
For $e_{\tr{Y}}=0.72$, a condition that $\chi_{\tr{tot}}^2/\tr{dof} \le
6.0/2=3.0$ yielded $0.548 \le z_{\tr{Y}} \le 0.983$. In the
best-fit model with $e_{\tr{Y}}=0.72$ and $z_{\tr{Y}}=0.958(0.548)$ 
(Table 1), the masses of Y
are $1.3(1.1)\times 10^9 \ms$ and $3.7(3.6)\times 10^9 \ms$ for regions in which the convergence contribution from Y satisfies
$\delta \kappa>\delta \kappa(\tr{A}2)$ and $\delta \kappa>\delta
\kappa(\tr{A}1)$\footnote{Each mass corresponds to a total mass within 
an elliptical aperture centered at the centroid
of Y.}, respectively and the one-dimensional velocity 
dispersion corresponding to $b_{\tr{Y}}$ 
is $\sigma_{\tr{Y}}\sim 25(28)\,\tr{km}\,\tr{s}^{-1}$. 

As shown in Figure 2, the convergence contribution $\delta \kappa $ from
Y is equal to or less than $\sim 5$ percent of the total convergence
$\kappa \sim 0.5$ at A1 and A2. $\delta \kappa(A2)$ is approximately
three times larger than $\delta \kappa(A1)$. 
The projected distance between the best-fit centroid of Y and 
A2 is 7.1\,kpc for $z_{\tr{Y}}=0.958$ and 5.3\,kpc for $z_{\tr{Y}}=0.548$.

In order to test the possibility that object Y is a sidelobe response to
A1 or A2, we simulated an ALMA observation using \ti{simalma} command in
CASA 4.5.3. We used the same antenna configurations and observation time
as those in the real observations. The value of the 
precipitable water vapor was adjusted to match the observed
noise in the final image. The source model was obtained from
the ALMA continuum image for a robust ($r=-1$) weighting using a
linear combination of multiple lensed images with a 
magnification weighting (see \citealt{inoue-minezaki2016} for its definition).
The total flux density was then adjusted to yield the observed value
for a natural and a robust ($r=-1$) weightings. 
We used the same $\ti{multi-scale}$ imaging 
as have been used for the corresponding real images. Here we
assumed $e_{\tr {Y}}=0.72$ and $z_{\tr{Y}}=0.548$.

As shown in Figure 3, the feature of the quadruply lensed spots and arcs 
was successfully reproduced though minor difference remains. 
However, we were not able to find any spots around
the position of Y. Therefore, it is unlikely that Y is related with
sidelobes of A1 or A2. The flux ratios
within an aperture radius of 0.21 arcsec were turned out to be 
$(\tr{A2}/\tr{A1}, \tr{B}/\tr{A1})
=(0.85,0.52)$ for a robust ($r=-1$) weighting, which are 
consistent with the observation of the dust continuum.
\subsection{Differential Extinction}
\begin{figure*}
\epsscale{0.7}
\hspace{-1cm}
\plotone{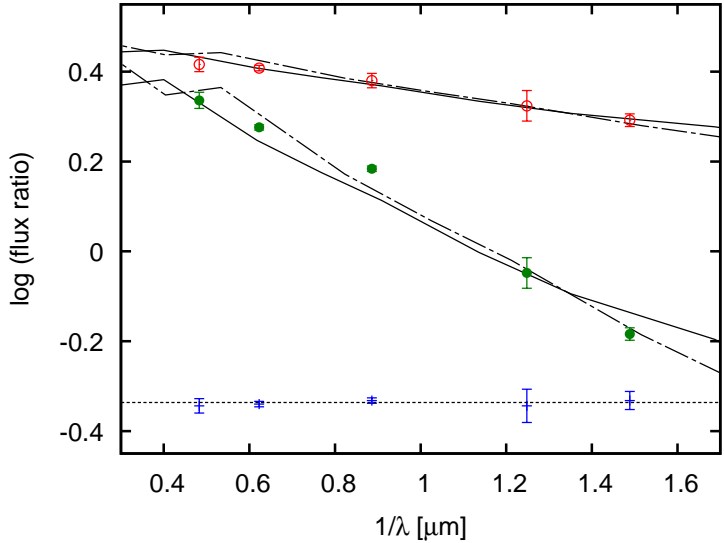}
\label{ff4}
 \caption{Optical and NIR flux ratios of MG\,0414+0534
 in different wavelengths.
 The open circles, the filled circles, and the plus marks represent
 the flux ratios of A1$/$B, A2$/$B, and C$/$B 
obtained from the CASTLES data archive, respectively \citep{falco1997}.
 The dot-dashed, and the solid lines represent
 the best-fit extinction curves for the flux ratios of A1$/$B and A2$/$B
 in which the source of extinction is assumed to reside 
at $z=0.5$, and $z=1.0$, respectively. The dotted line represents the
 weighted average of the flux ratio of C$/$B.}
\end{figure*}

From the unusually red colors, it has been speculated that 
MG\,0414+0534 is obscured by dust \citep[e.g.,][]{hewitt1992}. 
The optical and NIR flux ratios change 
with observing wavelength systematically,
which can be explained by the different amounts of extinction by
intervening dust in the line of sights.
However, the primary lensing galaxy G is a 
passively evolving early-type galaxy,
which is likely to be dust deficient \citep{tonry1999}.
Here we examine whether 
object Y can be the origin of the differential extinction.

Figure 4 shows the optical and near-infrared flux ratios
of A1$/$B, A2$/$B, and C$/$B in different wavelengths.
The flux data were taken from the CASTLES data archive \citep{falco1997}.
The F555W-band data were not used
because of their large photometric errors.
Clear reddening trends are found in the flux ratios of A1$/$B and A2$/$B
but not in C$/$B, which suggest that A1 and A2 are more obscured
by intervening dust in the line of sight than B and C,
as has been reported in previous studies.

Then we fit extinction curves to the flux ratios of 
A1$/$B and A2$/$B in different wavelengths 
to estimate the amount of differential extinction.
We adopted the dust extinction curve
for the Small Magellanic Cloud (SMC) bar \citep{gordon2003}
and assumed the MIR flux ratios as the intrinsic ones.
In order to examine the dust extinction from
a dwarf galaxy either in the intervening line of sight
or in the galaxy G, we 
put a hypothetical dust obscurer at $0.5$, and $1.0$
to estimate its amount of extinction, respectively.
In most cases, the reduced $\chi^2$ for the fits 
of extinction curves were larger than unity
if only the observational errors in the flux ratios are taken into account.
To achieve unity for the reduced $\chi^2$, we added the 
root-sum-square of systematic errors. 
The amount of systematic errors was $0.01$--$0.04$ dex.

In Figure 4, the best-fit extinction curves for A1$/$B and A2$/$B
are overlaid on the flux-ratio data.
The estimated differential extinctions of A1 relative to B
were $0.33\pm0.03$, $0.23\pm0.01$
and those of A2 relative to B were $1.11\pm0.05$, $0.77\pm0.04$
for $0.5$, $1.0$, respectively.
While the amount of extinction of the intervening obscurer
becomes larger when its redshift become smaller,
the flux ratios are fitted reasonably well
at any redshift assumed here.
We note that our estimates of the differential extinction
from an intervening obscurer at $z=1.0$
approximately agree with those in a 
previous study \citep{falco1999} in which the standard 
Galactic extinction law of $R_\tr{V}=3.1$ for $Y$ at 
$z_{\tr{d}}=z_\tr{L} = 0.958$ is assumed.

The weighted average of the flux ratios of C$/$B is also overlaid
in Figure 4. The estimated value $0.461\pm 0.002$ is consistent with
the C$/$B flux ratio in mid-infrared ($0.401\pm 0.043$) at $1.4$ sigmas,
and agrees well with those in our G$+$X$+$Y lens models ($0.469$ and
$0.459$) shown in Table 1.

We can estimate the ratio of the difference of convergence contribution $\delta \kappa$
between A2 and B to A1 and B as
$(\delta \kappa ({\rm A}2)-\delta \kappa ({\rm B}))/(\delta \kappa ({\rm
A}1)-\delta \kappa ({\rm B}))=4.3$ and $3.7$ for 
the redshifts of $z_{\rm Y}=0.548$ and $0.958$, respectively (Figure 2).
They are close to or slightly larger than the ratios of
the differential extinction of A2 to that of A1,
$\Delta A_{V}({\rm A}2)/\Delta A_{V}({\rm A}1)\approx 3.4$
for both the redshifts. Therefore, the large difference in the
differential extinction of A2 relative to A1 can be naturally explained
by our model in which object Y has an elongated SIE profile even though
A1 is located close to A2.

The dust mass column density $\Sigma_{\tr{dust}}$ in a sight
line per an amount of extinction is 
$\Sigma_{\tr{dust}}/A_V =
1.54\times 10^{-1}\,\ms \tr{pc}^{-2}/A_V$ for the Galaxy model and 
$2.66\times 10^{-1}\,\ms \tr{pc}^{-2}/A_V$ for the SMC dust model 
\citep{weingartner2001,draine2007}. Assuming that 
Y resides at $z_{\tr{Y}}=z_\tr{L}$, 
the dust mass column density
estimated from the amount of extinction 
in the site line of A1(A2) is 
$3.5(12) \times 10^{-2}\, \ms \tr{pc}^{-2}$ for the Galaxy dust model, and
$6.1(20) \times 10^{-2} \, \ms \tr{pc}^{-2}$ for the SMC
dust model. If the dust mass distribution traces the best-fit 
SIE mass profile of Y, then the estimated dust mass for the smallest elliptical
region centered at the centroid of Y that 
includes A1(A2) is $1.6(0.6) \times 10^7\,\ms$ 
for the Galaxy dust model, and $2.7(1.1) \times 10^7\,\ms$ 
for the SMC dust model. 
We find that the dust mass $\sim 10^7\,\ms$ 
estimated from the differential extinction 
does not significantly change if $z_{\tr{Y}}=0.5-1$ and it agrees with 
the value obtained from the sub-mm continuum emission of Y provided that the dust
temperature is $T_\tr{d}\sim 20 \,\tr{K}$, the emissivity index is 
$\beta=1-2$ and the redshift is $z_{\tr{Y}}=0.5-1$.

Thus these results indicate that object Y is a good candidate
for an intervening object that causes the differential extinction of A1 and A2.

\section{Discussions}
The detected 'object Y' is most probably a dusty
dark dwarf galaxy at a redshift of $0.5 \lesssim z_{\tr{Y}} \lesssim 1$. 
If Y is associated with a subhalo at $z_{\tr{Y}} \sim z_{\tr{L}}$, then it may be also 
associated with $\tr{H}_{\tr{I}}$ 21-cm absorption systems in the
lensing galaxy G, which has a large line-of-sight 
velocity $\sim 200\,\tr{km}\,\tr{s}^{-1}$ 
\citep{allison2016}.  The inferred $\tr{H}_{\tr{I}}$ column density of the
sum of three systems around $z=z_{\tr{L}}$ is $N_{\tr{H}_\tr{I}}=1.6\times 
10^{18}(T_{\tr{spin}}/f_{\tr{H}_{\tr{I}}})\,\tr{cm}^{-2}$
\citep{curran2007}, where $T_{\tr{spin}}$ is the spin
temperature and $f_{\tr{H}_{\tr{I}}}$ is the covering factor for the
background continuum source. If we assume $T_{\tr{spin}}=10^3\,\tr{K}$ 
and $f_{\tr{H}_{\tr{I}}}=0.5$, then the $\tr{H}_{\tr{I}}$ column density
estimated from absorption agrees with the value estimated from the 
difference in the hydrogen ($\tr{H}_{\tr{I}}+\tr{H}_2$) column densities in sight lines
to A1 and A2 relative to B, $\Delta N_{\tr{H}_\tr{I}}=(3.3 \pm
1.0)\times 10^{21}\,\tr{cm}^{-2}$, which was measured with the X-ray
spectra \citep{dai2006}. The latter can be converted into the sum of 
hydrogen mass column densities $\Sigma_\tr{H}$ in 
the sight lines to A1 and A2 as $\sim 2.6 \times 10^1\, \ms\tr{pc}^{-2}$.
Then the sum of baryon mass column densities at the sight lines 
is estimated as $\sim 3.7 \times 10^1\,\ms\tr{pc}^{-2}$
assuming the Solar abundance \citep{asplund2009}.
This value is comparable to the corresponding surface mass density $\sim 3.2 \times 10^1\,
\ms\tr{pc}^{-2}$ inferred from the best-fit SIE profile for Y. The 
hydrogen mass enclosing A1(A2) 
is then estimated as $\sim 3.1(1.1) \times 10^9\,\ms$ provided that the gas distribution traces
the best-fit SIE profile. This implies that 
the mass profile of the lens model is solely determined by baryon which
dominates over the dark matter. Such dark matter deficiency reminds us of 
'missing dark matter' in some dwarf galaxies \citep{oman2016}. 

The dust to gas ratio in the sight line to A1(A2) is
$A_V/N_\tr{H} \sim 2.6(3.2) \times 10^{-22} \,\tr{mag}\,
 \tr{cm}^2 \,\tr{H}^{-1}$ if the hydrogen column density in 
the sight line to A1(A2) is proportional to the surface mass density of
the best-fit SIE with $z_{\tr{Y}}=z_{\tr{L}}$. 
The dust to gas mass ratio in the sight line to A1(A2) is also estimated as
$\Sigma_\tr{dust}/\Sigma_{\tr{H}}\sim 0.005(0.006)$ for the Galaxy dust
model. These values are smaller than the Galactic values,
but larger than those for LMC and SMC \citep{weingartner2001}.
Conversely, if we assume the dust mass from extinction
and the dust-to-gas ratio for SMC,
the gas masses around A1 and A2 become much larger than
those inferred from the best-fit SIE mass profile assuming that baryon
traces dark matter. 
Therefore, we conclude that the dust to gas ratio and the dust
properties of Y are much similar to those in Galaxy than to those in LMC
or SMC if Y is associated with $\tr{H}_{\tr{I}}$ 
absorption systems at $z_{\tr{Y}} \sim z_\tr{L}$. 

If Y is associated with another dwarf galaxy in the line of sight, the most
probable redshift is $z_{\tr{Y}}=0.58$ at which the critical surface
density is the smallest. Since the critical surface density is 
a broad function of redshift, the reasonable range of 
redshift is also broad. For instance, for a 
redshift of $0.28 \le z_{\tr{Y}} \le 1.0$, the difference in the critical
surface density falls within 20 percent of the minimum
value. This is consistent with our lensing constraint: $0.548 \le 
z_{\tr{Y}} \le 0.958$. Indeed, a recent theoretical analysis predicts that the expected 
convergence contribution from intervening structures is much 
larger than that from subhalos in MG\,0414+0534 \citep{inoue2016}.

Although object Y seems to have a relatively
large size $\gtrsim 5$\,kpc for a galaxy with one-dimensional 
velocity dispersion of $\sim 30$\,km/s, 
the optical/NIR emission 
seems to be extremely small as $\lesssim
0.06\,\mu$Jy at the I band\footnote{The upper limit corresponds to
3\,$\sigma$ error in the I band flux \citep{falco1997}.}. If Y is at $z=0.5$, the corresponding 
stellar mass would be $M_\star \lesssim 10^6\,\ms$ if the SED is similar to 
that of faint submillimeter galaxies like LESS 34 \citep{wiklind2014}. 
However, it is unnatural unless huge amount of
dust is hiding luminous stellar components. 
Perhaps, Y is an object like recently 
discovered ultra diffuse galaxies (UDGs) in Coma 
\citep{van-dokkum2015, koda2015}. UDGs have
stellar masses typical of dwarf galaxies but effective radii of
Milky-Way sized objects. The stellar components may have been expanded due to 
outflow caused by starburst \citep{cintio2016} or AGN activity. Then the
surface brightness in the optical/NIR band can be extremely small
despite being gas-rich. The required high ellipticity $e_\tr{Y}\sim 0.7$
for Y is consistent with the shape of UDGs if prolate \citep{burkert2016} and
that of dark matter halos that host the 
Local Group dwarf spheroidal galaxies if oblate \citep{hayashi2015}.
More detailed analysis of Y would give us a clue about the
origin of past starburst activity. If Y is turned out to be
residing in the intervening line of sight, then a substantial portion of
faint submillimeter galaxies in the universe may be attributed to such
gas-rich dusty dark dwarf galaxies.
\acknowledgments
KTI acknowledges support from the ALMA Japan Research Grant
of NAOJ Chile Observatory, NAOJ-ALMA-0110. SM is 
supported by the Ministry of Science and Technology (MoST) 
of Taiwan, MoST 103-2112-M-001-032-MY3. MC acknowledges 
support from MEXT Grant-in-Aid for Scientific Research on Innovative
Areas (No. 15H05889, 16H01086). This paper makes use of the following ALMA data:
ADS/JAO.ALMA$\#$2013.1.01110.S. ALMA is a partnership of ESO (representing its member states), NSF (USA) and NINS (Japan), together with NRC (Canada) 
and NSC and ASIAA (Taiwan), and KASI (Republic of Korea) in
cooperation with the Republic of Chile. The Joint ALMA Observatory is
operated by ESO, AUI/NRAO and NAOJ.
%


\begin{thebibliography}{}
\expandafter\ifx\csname natexlab\endcsname\relax\def\natexlab#1{#1}\fi
\providecommand{\url}[1]{\href{#1}{#1}}

\bibitem[{{Allison} {et~al.}(2016){Allison}, {Moss}, {Macquart}, {Curran},
  {Duchesne}, {Mahony}, {Sadler}, {Whiting}, {Bannister}, {Chippendale},
  {Edwards}, {Harvey-Smith}, {Heywood}, {Indermuehle}, {Lenc}, {Marvil},
  {McConnell}, \& {Sault}}]{allison2016}
{Allison}, J.~R., {Moss}, V.~A., {Macquart}, J.-P., {et~al.} 2016, ArXiv
  e-prints, arXiv:1611.00863

\bibitem[{{Asplund} {et~al.}(2009){Asplund}, {Grevesse}, {Sauval}, \&
  {Scott}}]{asplund2009}
{Asplund}, M., {Grevesse}, N., {Sauval}, A.~J., \& {Scott}, P. 2009, \araa, 47,
  481

\bibitem[{{Barvainis} \& {Ivison}(2002)}]{barvainis2002}
{Barvainis}, R., \& {Ivison}, R. 2002, \apj, 571, 712

\bibitem[{{Burkert}(2016)}]{burkert2016}
{Burkert}, A. 2016, ArXiv e-prints, arXiv:1609.00052

\bibitem[{{Chiba}(2002)}]{chiba2002}
{Chiba}, M. 2002, \apj, 565, 17

\bibitem[{Chiba {et~al.}(2005)Chiba, Minezaki, Kashikawa, Kataza, \&
  Inoue}]{chiba2005}
Chiba, M., Minezaki, T., Kashikawa, N., Kataza, H., \& Inoue, K.~T. 2005, \apj,
  627, 53

\bibitem[{{Cornwell}(2008)}]{cornwell2008}
{Cornwell}, T.~J. 2008, IEEE Journal of Selected Topics in Signal Processing,
  2, 793

\bibitem[{{Curran} {et~al.}(2007){Curran}, {Darling}, {Bolatto}, {Whiting},
  {Bignell}, \& {Webb}}]{curran2007}
{Curran}, S.~J., {Darling}, J., {Bolatto}, A.~D., {et~al.} 2007, \mnras, 382,
  L11

\bibitem[{{Dai} {et~al.}(2006){Dai}, {Kochanek}, {Chartas}, \&
  {Mathur}}]{dai2006}
{Dai}, X., {Kochanek}, C.~S., {Chartas}, G., \& {Mathur}, S. 2006, \apj, 637,
  53

\bibitem[{{Dalal} \& {Kochanek}(2002)}]{dalal-kochanek2002}
{Dalal}, N., \& {Kochanek}, C.~S. 2002, \apj, 572, 25

\bibitem[{{Di Cintio} {et~al.}(2016){Di Cintio}, {Brook}, {Dutton},
  {Macci{\`o}}, {Obreja}, \& {Dekel}}]{cintio2016}
{Di Cintio}, A., {Brook}, C.~B., {Dutton}, A.~A., {et~al.} 2016, ArXiv
  e-prints, arXiv:1608.01327

\bibitem[{{Draine} \& {Li}(2007)}]{draine2007}
{Draine}, B.~T., \& {Li}, A. 2007, \apj, 657, 810

\bibitem[{{Dunne} {et~al.}(2000){Dunne}, {Eales}, {Edmunds}, {Ivison},
  {Alexander}, \& {Clements}}]{dunne2000}
{Dunne}, L., {Eales}, S., {Edmunds}, M., {et~al.} 2000, \mnras, 315, 115

\bibitem[{{Falco} {et~al.}(1997){Falco}, {Lehar}, \& {Shapiro}}]{falco1997}
{Falco}, E.~E., {Lehar}, J., \& {Shapiro}, I.~I. 1997, \apj, 113, 540

\bibitem[{{Falco} {et~al.}(1999){Falco}, {Impey}, {Kochanek}, {Leh{\'a}r},
  {McLeod}, {Rix}, {Keeton}, {Mu{\~n}oz}, \& {Peng}}]{falco1999}
{Falco}, E.~E., {Impey}, C.~D., {Kochanek}, C.~S., {et~al.} 1999, \apj, 523,
  617

\bibitem[{{Gordon} {et~al.}(2003){Gordon}, {Clayton}, {Misselt}, {Landolt}, \&
  {Wolff}}]{gordon2003}
{Gordon}, K.~D., {Clayton}, G.~C., {Misselt}, K.~A., {Landolt}, A.~U., \&
  {Wolff}, M.~J. 2003, \apj, 594, 279

\bibitem[{{Hayashi} \& {Chiba}(2015)}]{hayashi2015}
{Hayashi}, K., \& {Chiba}, M. 2015, \apj, 810, 22

\bibitem[{{Hewitt} {et~al.}(1992){Hewitt}, {Turner}, {Lawrence}, {Schneider},
  \& {Brody}}]{hewitt1992}
{Hewitt}, J.~N., {Turner}, E.~L., {Lawrence}, C.~R., {Schneider}, D.~P., \&
  {Brody}, J.~P. 1992, \apj, 104, 968

\bibitem[{{Hezaveh} {et~al.}(2016){Hezaveh}, {Dalal}, {Marrone}, {Mao},
  {Morningstar}, {Wen}, {Blandford}, {Carlstrom}, {Fassnacht}, {Holder},
  {Kemball}, {Marshall}, {Murray}, {Perreault Levasseur}, {Vieira}, \&
  {Wechsler}}]{hezaveh2016}
{Hezaveh}, Y.~D., {Dalal}, N., {Marrone}, D.~P., {et~al.} 2016, \apj, 823, 37

\bibitem[{{Inoue}(2015)}]{inoue2015}
{Inoue}, K.~T. 2015, \mnras, 447, 1452

\bibitem[{{Inoue}(2016)}]{inoue2016}
---. 2016, \mnras, 461, 164

\bibitem[{{Inoue} \& {Chiba}(2005{\natexlab{a}})}]{inoue2005a}
{Inoue}, K.~T., \& {Chiba}, M. 2005{\natexlab{a}}, \apj, 634, 77

\bibitem[{{Inoue} \& {Chiba}(2005{\natexlab{b}})}]{inoue2005b}
---. 2005{\natexlab{b}}, \apj, 633, 23

\bibitem[{{Inoue} {et~al.}(2016){Inoue}, {Minezaki}, {Matsushita}, \&
  {Chiba}}]{inoue-minezaki2016}
{Inoue}, K.~T., {Minezaki}, T., {Matsushita}, S., \& {Chiba}, M. 2016, \mnras,
  457, 2936

\bibitem[{{Inoue} \& {Takahashi}(2012)}]{inoue-takahashi2012}
{Inoue}, K.~T., \& {Takahashi}, R. 2012, \mnras, 426, 2978

\bibitem[{Kochanek \& Dalal(2004)}]{kochanek2004}
Kochanek, C.~S., \& Dalal, N. 2004, \apj, 610, 69

\bibitem[{{Koda} {et~al.}(2015){Koda}, {Yagi}, {Yamanoi}, \&
  {Komiyama}}]{koda2015}
{Koda}, J., {Yagi}, M., {Yamanoi}, H., \& {Komiyama}, Y. 2015, \apjl, 807, L2

\bibitem[{{Koopmans}(2005)}]{koopmans2005}
{Koopmans}, L.~V.~E. 2005, \mnras, 363, 1136

\bibitem[{{Kormann} {et~al.}(1994){Kormann}, {Schneider}, \&
  {Bartelmann}}]{kormann1994}
{Kormann}, R., {Schneider}, P., \& {Bartelmann}, M. 1994, Astronomy and
  Astrophysics, 284, 285

\bibitem[{Lawrence {et~al.}(1995)Lawrence, Elston, Januzzi, \&
  Turner}]{lawrence1995}
Lawrence, C.~R., Elston, R., Januzzi, B.~T., \& Turner, E.~L. 1995,
  Astronomical Journal, 110, 2570

\bibitem[{{MacLeod} {et~al.}(2013){MacLeod}, {Jones}, {Agol}, \&
  {Kochanek}}]{macleod2013}
{MacLeod}, C.~L., {Jones}, R., {Agol}, E., \& {Kochanek}, C.~S. 2013, \apj,
  773, 35

\bibitem[{{Mao} \& {Schneider}(1998)}]{mao1998}
{Mao}, S., \& {Schneider}, P. 1998, \mnras, 295, 587

\bibitem[{{McMullin} {et~al.}(2007){McMullin}, {Waters}, {Schiebel}, {Young},
  \& {Golap}}]{mcmullin2007}
{McMullin}, J.~P., {Waters}, B., {Schiebel}, D., {Young}, W., \& {Golap}, K.
  2007, in Astronomical Society of the Pacific Conference Series, Vol. 376,
  Astronomical Data Analysis Software and Systems XVI, ed. R.~A. {Shaw},
  F.~{Hill}, \& D.~J. {Bell}, 127

\bibitem[{{Metcalf}(2005)}]{metcalf2005a}
{Metcalf}, R.~B. 2005, \apj, 629, 673

\bibitem[{{Metcalf} {et~al.}(2004){Metcalf}, {Moustakas}, {Bunker}, \&
  {Parry}}]{metcalf2004}
{Metcalf}, R.~B., {Moustakas}, L.~A., {Bunker}, A.~J., \& {Parry}, I.~R. 2004,
  \apj, 607, 43

\bibitem[{Metcalf \& Zhao(2002)}]{metcalf2002}
Metcalf, R.~B., \& Zhao, H. 2002, \apj Letters, 567, L5

\bibitem[{{Minezaki} {et~al.}(2009){Minezaki}, {Chiba}, {Kashikawa}, {Inoue},
  \& {Kataza}}]{minezaki2009}
{Minezaki}, T., {Chiba}, M., {Kashikawa}, N., {Inoue}, K.~T., \& {Kataza}, H.
  2009, \apj, 697, 610

\bibitem[{{More} {et~al.}(2009){More}, {McKean}, {More}, {Porcas}, {Koopmans},
  \& {Garrett}}]{more2009}
{More}, A., {McKean}, J.~P., {More}, S., {et~al.} 2009, \mnras, 394, 174

\bibitem[{{Oman} {et~al.}(2016){Oman}, {Navarro}, {Sales}, {Fattahi}, {Frenk},
  {Sawala}, {Schaller}, \& {White}}]{oman2016}
{Oman}, K.~A., {Navarro}, J.~F., {Sales}, L.~V., {et~al.} 2016, \mnras, 460,
  3610

\bibitem[{{Planck Collaboration} {et~al.}(2016){Planck Collaboration}, {Ade},
  {Aghanim}, {Arnaud}, {Ashdown}, {Aumont}, {Baccigalupi}, {Banday},
  {Barreiro}, {Bartlett}, \& et~al.}]{planck2016}
{Planck Collaboration}, {Ade}, P.~A.~R., {Aghanim}, N., {et~al.} 2016, \aap,
  594, A13

\bibitem[{Ros {et~al.}(2000)Ros, Guirado, Marcaide, Perez-Torres, Falco, Munoz,
  Alberdi, \& Lara}]{ros2000}
Ros, E., Guirado, J.~C., Marcaide, J.~M., {et~al.} 2000, Astronomy and
  Astrophysics, 362, 845

\bibitem[{Schechter \& Moore(1993)}]{schechter1993}
Schechter, P.~L., \& Moore, C.~B. 1993, Astronomical Journal, 105, 1

\bibitem[{{Sugai} {et~al.}(2007){Sugai}, {Kawai}, {Shimono}, {Hattori},
  {Kosugi}, {Kashikawa}, {Inoue}, \& {Chiba}}]{sugai2007}
{Sugai}, H., {Kawai}, A., {Shimono}, A., {et~al.} 2007, \apj, 660, 1016

\bibitem[{{Takahashi} \& {Inoue}(2014)}]{takahashi-inoue2014}
{Takahashi}, R., \& {Inoue}, K.~T. 2014, \mnras, 440, 870

\bibitem[{Tonry \& Kochanek(1999)}]{tonry1999}
Tonry, J.~L., \& Kochanek, C.~S. 1999, Astronomical Journal, 117, 2034

\bibitem[{{Trotter} {et~al.}(2000){Trotter}, {Winn}, \& {Hewitt}}]{trotter2000}
{Trotter}, C.~S., {Winn}, J.~N., \& {Hewitt}, J.~N. 2000, \apj, 535, 671

\bibitem[{{van Dokkum} {et~al.}(2015){van Dokkum}, {Abraham}, {Merritt},
  {Zhang}, {Geha}, \& {Conroy}}]{van-dokkum2015}
{van Dokkum}, P.~G., {Abraham}, R., {Merritt}, A., {et~al.} 2015, \apjl, 798,
  L45

\bibitem[{{Vegetti} {et~al.}(2014){Vegetti}, {Koopmans}, {Auger}, {Treu}, \&
  {Bolton}}]{vegetti2014}
{Vegetti}, S., {Koopmans}, L.~V.~E., {Auger}, M.~W., {Treu}, T., \& {Bolton},
  A.~S. 2014, \mnras, 442, 2017

\bibitem[{{Vegetti} {et~al.}(2012){Vegetti}, {Lagattuta}, {McKean}, {Auger},
  {Fassnacht}, \& {Koopmans}}]{vegetti2012}
{Vegetti}, S., {Lagattuta}, D.~J., {McKean}, J.~P., {et~al.} 2012, Nature, 481,
  341

\bibitem[{{Weingartner} \& {Draine}(2001)}]{weingartner2001}
{Weingartner}, J.~C., \& {Draine}, B.~T. 2001, \apj, 548, 296

\bibitem[{{Wiklind} {et~al.}(2014){Wiklind}, {Conselice}, {Dahlen},
  {Dickinson}, {Ferguson}, {Grogin}, {Guo}, {Koekemoer}, {Mobasher},
  {Mortlock}, {Fontana}, {Dav{\'e}}, {Yan}, {Acquaviva}, {Ashby}, {Barro},
  {Caputi}, {Castellano}, {Dekel}, {Donley}, {Fazio}, {Giavalisco}, {Grazian},
  {Hathi}, {Kurczynski}, {Lu}, {McGrath}, {de Mello}, {Peth}, {Safarzadeh},
  {Stefanon}, \& {Targett}}]{wiklind2014}
{Wiklind}, T., {Conselice}, C.~J., {Dahlen}, T., {et~al.} 2014, \apj, 785, 111

\bibitem[{{Xu} {et~al.}(2012){Xu}, {Mao}, {Cooper}, {Gao}, {Frenk}, {Angulo},
  \& {Helly}}]{xu2012}
{Xu}, D.~D., {Mao}, S., {Cooper}, A.~P., {et~al.} 2012, \mnras, 421, 2553

\bibitem[{{Xu} {et~al.}(2010){Xu}, {Mao}, {Cooper}, {Wang}, {Gao}, {Frenk}, \&
  {Springel}}]{xu2010}
---. 2010, \mnras, 408, 1721

\bibitem[{{Xu} {et~al.}(2009){Xu}, {Mao}, {Wang}, {Springel}, {Gao}, {White},
  {Frenk}, {Jenkins}, {Li}, \& {Navarro}}]{xu2009}
{Xu}, D.~D., {Mao}, S., {Wang}, J., {et~al.} 2009, \mnras, 398, 1235

\end{thebibliography}

\end{document}